\documentclass[fleqn,twoside,twocolumn,nofootinbib,showkeys]{revtex4} 
\usepackage[sec,nocpr]{ujp} 

\begin{document}
\title[Self-Organization and Chaos
 in the Metabolism of Hemostasis]
{SELF-ORGANIZATION AND CHAOS\\
 IN THE METABOLISM OF HEMOSTASIS\\ IN A BLOOD VESSEL}%
\author{V.I. Grytsay}
\affiliation{\bitp}
\address{\bitpaddr}
\email{vgrytsay@bitp.kiev.ua}

\udk{577.3} \pacs{05.45.-a, 05.45.Pq,\\[-3pt]
05.65.+b} \razd{\secx}

\autorcol{V.I.\hspace*{0.7mm}Grytsay}

\setcounter{page}{1}%

\begin{abstract}
A mathematical model of the metabolic process of formation of the
hemostasis in a blood-carrying vessel is constructed. As distinct
from the earlier developed model of the multienzyme
prostacyclin-thromboxane system of blood, this model includes, for
the first time, the influence of the level of ``bad cholesterol'',
i.e., low-density lipoproteins (LDLs), on the hemostasis. The
conditions, under which the self-organization of the system appears,
and the modes of autooscillations and chaos in the metabolic
process, which affects the formation of hemostasis and the
development of thrombophilia, are found. With the aid of a
phase-parametric diagram, the scenario of their appearance is
studied. The bifurcations of doubling of a period and the transition
to chaotic oscillations as a result of the intermittence are found.
The obtained strange attractors are formed due to a mixing funnel.
The full spectra of Lyapunov's indices, KS-entropies,
``predictability horizons'', and Lyapunov's dimensions of strange
attractors are calculated. The reasons for a change in the cyclicity
of the given metabolic process, its stability, and the physiological
manifestation in the blood-carrying system are discussed. The role
of physiologically active substances in a decrease in the level of
cholesterol in blood vessels is estimated.
\end{abstract}

\keywords{hemostasis, LDLs, self-organization, strange attractor,
phase-parametric diagram, Lyapunov's indices, KS-entropy.}

\maketitle

\noindent Multicellular organism is a holistic system, whose cells
are specialized for the execution of various functions. Their
structural-functional connections form the self-organization in the
system. The interaction inside organism is realized via complex
regulating, coordinating, and correcting mechanisms with
participation of neural, humoral, exchanging, and other processes.
Many separate mechanisms, which regulate internal cellular and
intercellular mutual relations, form opposite (antagonistic)
actions, by balancing one another. This leads to the establishment
of a variable physiological equilibrium in organism, which allows an
alive system to hold a relative dynamical stationary state, despite
the changes in the environment and the variations arising in the
process of vital activity of organism. From the viewpoint of
biophysics, it is a state, in which all processes responsible for
the energetic transformations in organism are in a dynamical
equilibrium. Such state is called the homeostasis.

The blood composition steadiness has a special meaning for the vital
activity of organism. Two antagonistic systems exist in blood
vessels: those responsible for the coagulation and anticoagulation
of blood. Due to their presence, the functional equilibrium, which
hampers the coagulation of blood inside vessels, is formed. This
state is called the hemostasis \cite{1,2,3,4}.

But if the wall of a vessel is damaged, the blood flow decelerates,
or the functional state of the system of hemostasis is changed, then
the conditions for the development of thrombophilia and thrombosis
\mbox{arise.}\looseness=1

\begin{figure}%
\vskip1mm
\includegraphics[width=\column]{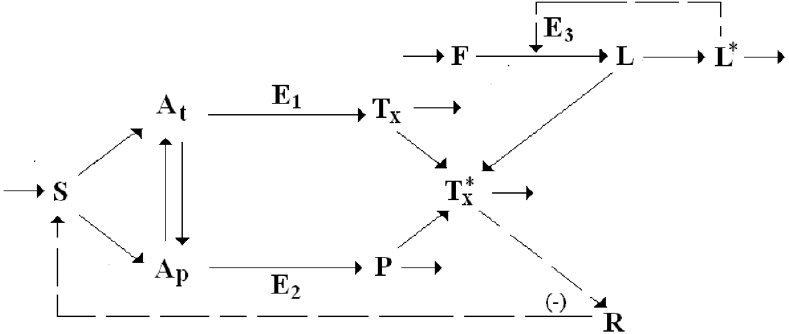}
\vskip-3mm\caption{General kinetic diagram of the hemostasis of a
blood vessel}\label{fig:1}
\end{figure}

The regulators of the coagulation of blood are thromboxanes and
prostacyclins. Thromboxanes and prostacyclins are created, respectively, in
thrombocytes and in endothelial internal cells of a vessel. The high level
of thromboxanes favors the aggregation of thrombocytes, which leads to the
coagulation of blood. On the contrary, prostacyclins hamper the coagulation
of blood, being the inhibitors of the aggregation of thrombocytes. The
establisnment of a dynamical equilibrium between those systems leads to the
establishment of a hemostasis.

A number of researchers modeled this and other analogous processes
\cite{5,8}. Those calculations were partially carried out with
participation of the author. Later with the use of those results, a
mathematical model of a multienzyme prostacyclin-thromboxane system
of blood was developed in \cite{9,10,11,12}. With the use of the
theory of dissipative structures, the conditions of
self-organization of the system and the appearance of stationary
stable autoperiodic oscillations were found. Moreover, the
conditions of appearance of hemophilia and thrombosis and the
physiological influence of autooscillatory modes on the
blood-carrying system were studied.

In the present work, we continue the study of this system with regard for
the presence of cholesterol in blood.

One of the basic factors affecting the functional state of
hemostasis is the level of ``bad cholesterol'' in blood. It is
determined by the level of low-density lipoproteins (LDLs). At its
excess concentration, atherosclerosis of vessels develops
\cite{13,14,15,16}.

The excess of these particles penetrates into the arterial wall, by
accumulating in it. In this case, their chemical formula is changed.
Modified LDLs stimulate endothelial cells to the activation of adhesion
receptors, which join monocytes of blood and T-cells. In addition,
endothelial cells give off a secretion, namely, chemokines, which entice
monocytes in a trap of intima. In intima, monocytes ripen and become active
macrophages. They absorb LDLs, by filling themselves with drops of fat.
These foamy fat-loaded macrophages and T-cells form fat strips, which are
predecessors of complex arterial patches, which disfigure arteries later on.
Molecules from inflammation can favor the further growth of a patch and the
formation of a fibrous cap above the lipid core, which causes the stenosis
of a vessel. The further growth of a patch can cause its rupture and the
appearance of thrombosis.

We will study the influence of LDLs on the self-organization of the
metabolic process of hemostasis in blood vessels and on the
appearance of thrombophilia. The role of drugs for the purification
of blood vessels from cholesterol will be evaluated.\vspace*{-2mm}

\section{Mathematical Model}

We will construct a mathematical model according to the general
diagram of the hemostasis of a blood vessel, which is presented in
Fig.~\ref{fig:1}.

This system is open and nonequilibrium. For it, the input substance
is irreplaceable arachidonic fat acid $S$, which enters blood from
digestive tract. Under the action of phospholipases, it is
accumulated in thrombocytes $A_t $ and endothelial cells $A_p $.
Then it is transformed by prostaglandin-$H$-synthetase of
thrombocytes $E_1 $ and prostaglandin-$H$-synthetase of
prostacyclins $E_2 $, by forming, respectively, thromboxanes $T_x $
and prostacyclins $P$. Their levels depend on the activity of a
dissipative outflow of either component. The output product of the
system is a complex of the mentioned quantities formed at the
aggregation of thrombocytes $T_x^\ast $. The regulation of the
processes of hemostasis and thrombus-forming by prostaglandin is
determined by the concentration $R$ of the regulating component,
namely cyclic adenosine monophosphate (cAMP). This happens under the
action of a negative feedback, which affects the level of activity
of phospholipases of thrombocytes $A_t $ and endothelial cells
$A_p$. As its value varies, the different modes are formed.

The model involves the exchange of arachidonic acid between
thrombocytes and endothelial cells ($A_t \leftrightarrow A_p )$,
inactivation of enzymes at proteolysis, and the consumption of
components in other \mbox{processes.}

The second input substance of the system is fat molecules $F$. They
are transported by blood of arteries and affect the level of ``bad
cholesterol'', LDLs. Its level is given by the variable $L$. LDLs
are formed in liver and small intestine. In increased
concentrations, LDLs influence the aggregation of thrombocytes
$T_x^\ast $ and are accumulated in the walls of arteries, by forming
oxidized lipoproteins $L^\ast $. The antithrombic system partially
decomposes them. As a result of the deposition of ``bad
cholesterol'', foamy fat-loaded macrophages are formed, which
increases, in turn, the level of LDLs in artery. In the metabolic
process, the positive feedback regulated by enzyme $E_3 $ is formed.
The accumulation of cholesterol in artery and the growth of patches
cause thrombophilia. In this case, the blood-carrying channel of
artery becomes narrower, and stenosis arises. Depending on the
dissipation of LDLs, autooscillatory and chaotic modes can appear in
the metabolic process of hemostasis of a blood vessel instead of
stationary ones.

The mathematical model of this process is presented by
Eqs.\,(\ref{eq1})--(\ref{eq12}). This system includes 12
differential equations describing changes in the concentrations of
components according to the general scheme given in
Fig.~\ref{fig:1}. In the construction of the model, the law of mass
action and the kinetics of enzyme catalysis are used. The equations
involve the balance of masses for the intermediate products of
reactions between separate stages of the metabolic process:
\[
\frac{dA_t }{dt}=\frac{k_5 S}{(1+S+R^2)(1+k_6 T_x )}\,-
\]\vspace*{-7mm}
\[-\,\frac{k_7 A_t E_1 }{(1+A_t +k_1 T_x )(1+E_1 )}\,+
\]\vspace*{-7mm}
\begin{equation}
\label{eq1} +\,k_p A_p -k_t A_t -\alpha _1 A_t ,
\end{equation}\vspace*{-7mm}
\begin{equation}
\label{eq2} \frac{dT_x }{dt}=\frac{k_7 A_t E_1 }{(1+A_t +k_1 T_x
)(1+E_1 )}-\frac{k_8 T_x^4 }{(k_9 +T_x^4 )}-\alpha _2 T_x ,
\end{equation}\vspace*{-7mm}
\[
\frac{dA_p }{dt}=\frac{k_2 SR^2}{(1+S+k_3 A_p )(k_4
\!+R^2)}-\frac{k_{10} A_p E_2 }{(1+\!A_p )(1+\!E_2 )}\,+
\]\vspace*{-7mm}
\begin{equation}
\label{eq3} +\,k_t A_t -k_p A_p -\alpha _3 A_p ,
\end{equation}
\begin{equation}
\label{eq4} \frac{dP}{dt}=\frac{k_{10} A_p E_2 }{(1+A_p )(1+E_2
)}-\frac{k_{11} T_x^\ast P^4}{(1+T_x^\ast )(k_{12} +P^4)}-\alpha _4
P,
\end{equation}\vspace*{-9mm}
\[
\frac{dE_1 }{dt}=\frac{k_{13} A_T }{(1+A_T )(1+R^4)}\,-
\]\vspace*{-7mm}
\begin{equation}
\label{eq5} -\,\frac{k_7 A_t E_1 }{(1+A_t +k_1 T_x )(1+E_1 )}-\alpha
_5 E_1 ,
\end{equation}\vspace*{-7mm}
\[
\frac{dE_2 }{dt}=\frac{k_{15} A_p T_x^{\ast 4} }{(k_{16} +A_p
)(k_{17} +T_x^{\ast 4} )}\,-
\]\vspace*{-7mm}
\begin{equation}
\label{eq6} -\,\frac{k_{10} A_p E_2 }{(1+A_p )(1+E_2 )}-\alpha _6
E_2 ,
\end{equation}\vspace*{-7mm}
\begin{equation}
\label{eq7} \frac{dR}{dt}=k_{18} \frac{k_{19} +T_x^{\ast 4} }{k_{20}
+(T_x^\ast +k_{21} R)^4}-\alpha _7 R,
\end{equation}\vspace*{-7mm}
\begin{equation}
\label{eq8} \frac{dT_x^\ast }{dt}=k_8 \frac{L+T_x^4 }{k_9 +L+T_x^4
}-\frac{k_{11} T_x^\ast P^4}{(1+T_x^\ast )(k_{12} +P^4)}-\alpha _8
T_x^\ast\!.
\end{equation}\vspace*{-9mm}
\begin{equation}
\label{eq9} \frac{dF}{dt}=F_0 -l\frac{E_3 }{1+E_3 }\,
\frac{F}{1+F+L},
\end{equation}\vspace*{-7mm}
\begin{equation}
\label{eq10} \frac{dL}{dt}=k\frac{E_3 }{1+E_3 }\,
\frac{F}{1+F+L}-\mu \frac{LL^\ast }{1+L+L^\ast },
\end{equation}\vspace*{-7mm}
\begin{equation}
\label{eq11} \frac{dL^\ast }{dt}=\mu _1 \frac{LL^\ast }{1+L+L^\ast
}-\mu _0 L^\ast\!,
\end{equation}\vspace*{-7mm}
\begin{equation}
\label{eq12} \frac{dE_3 }{dt}=E_{3_0 } L^\ast \frac{F}{1+F}\,
\frac{N}{N+L}-\alpha _9 E_3 .
\end{equation}
The model is specified by the following collection of parameters:
$k=4;$ $k_1 =3;$ $k_2 =1;$ $k_3 =5;$ $k_4 =$ $=10;$ $k_5 =2.1;$ $k_6
=5; $ $k_7 =2;$ $k_8 =1.5;$ $k_9 =5;$ $k_{10} =0.75;$ $k_{11} =0.3;
$ $k_{12} =15;$ $k_{13} =0.75;$ $k_{15} =$ $=1;$ $k_{16} =0.5;$ $
k_{17} =5;$ $k_{18} =5;$ $k_{19} =0.02;$ $k_{20} =$ $=25;$ $k_{21}
=0.5; $ $k_p =0.1;$ $k_t =0.1;$ $S=2;$ $\alpha _1 =$ $=0.01;$
$\alpha _2 =0.01;$ $\alpha _3 =0.01;$ $\alpha _4 =0.173;$ $\alpha _5
=$ $=0.05; $ $\alpha _6 =0.07;$ $\alpha _7 =0.2;$ $\alpha _8
=0.0021;$ $ \alpha _9 =0.2;$ $F_0 =0.01;$ $l=2;$ $\mu =4;$ $\mu _0
=0.42;$ $ \mu _1 =2.3;$ $E_{3_0 } =$ $=11;$ $N=0.05 $. The study of
solutions of the given mathematical model (\ref{eq1})--(\ref{eq12})
was performed with the aid of the theory of nonlinear differential
equations \cite{17,18} and the methods of modeling of biochemical
systems, which were applied by the author in
\cite{19,20,21,22,23,24,25,26,27,28,29,30,31,32,33,34,35,36,37,38,39,40}.
In numerical calculations, the Runge--Kutta--Merson method was used.
The accuracy of calculations is about $10^{-8}$.

The other mathematical models of biochemical processes can be found,
for example, in \cite{41,42,43,44}.

\begin{figure*}%
\vskip1mm
\includegraphics[width=12cm]{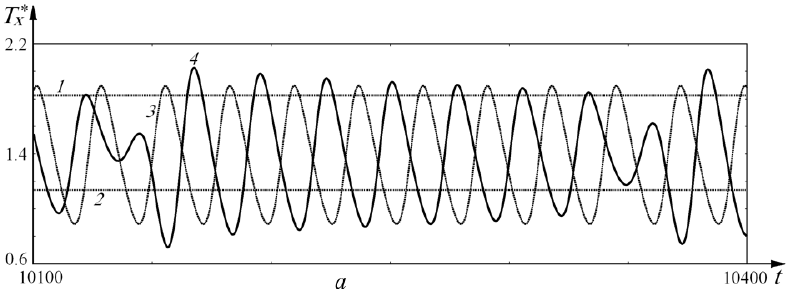}\\[2mm]
\includegraphics[width=12cm]{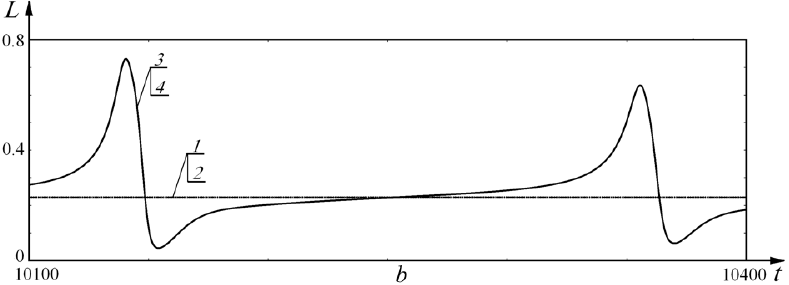}
\vskip-2mm\parbox{12.0cm}{\caption{Kinetic curves of the stationary
mode for $\alpha _7 = 0.005$ and $\mu_0=0.42$ (\textit{1}),
stationary mode for $\alpha _7 = 1.0$ and $\mu_0=0.42$ (\textit{2}),
autoperiodic mode for $\alpha _7 = 0.2$ and $\mu _0 = 0.42$
(\textit{3}), and a chaotic mode for $\alpha _7 = 0.2$ and $\mu _0 =
0.437$ (\textit{4}) on planes: ($T_x^\ast, t $) (\textit{a}); ($L,
t$) (\textit{b})\label{fig:2}}}\vspace*{1.5mm}
\end{figure*}

\section{Results of Studies}\vspace*{-1mm}

The presented mathematical model is given by the system of nonlinear
differential equations (\ref{eq1})--(\ref{eq12}), which describes
the open nonlinear biochemical system of the hemostasis of a blood
vessel in the presence of cholesterol in blood.

The study of the dynamics of the metabolic process of
thrombosis-antithrombosis was carried out without cholesterol in
blood in works \cite{9,10,11,12}. Here, we calculated the modes
depending on the value of the parameter of $\alpha _7 $
characterizing the dissipation of cAMP $(R)$. Like the previous
works, the system including cholesterol possesses two stationary
modes. In Fig.~\ref{fig:2},~\textit{a} we show the dependence of the
kinetics of $T_x^\ast $ on the parameter $\alpha _7 $. For large
($\alpha _7 = 1$) and small ($\alpha _7 = 0.005$) values of this
parameter, stationary modes are established. For $\alpha _7 = 1$, a
thermodynamical branch (\textit{1}) is formed, and, for $\alpha _7 =
0.005$, we observe the stationary dissipative structure
(\textit{2}). In the interval between these values for $\alpha _7 =
0.2$, the systems of thrombosis-antithrombosis are characterized by
the autoperiodic mode (\textit{3}). In other words, for the
parameter of dissipation of cholesterol $\mu _0 = 0.439$, the
kinetics of the process is invariable. But if $\mu _0 = 0.437$, the
autoperiodic mode is replaced by the chaotic mode (\textit{4}).
Thus, the kinetics is determined by the interaction of the systems
of thrombosis-antithrombosis and the rate of removal of cholesterol
from a blood vessel. In Fig.~\ref{fig:2},~\textit{b}, we show how
the level of LDLs is changed, by depending on the type of a mode. In
the stationary modes of hemostasis (\textit{1}) and (\textit{2}),
the level of ``bad cholesterol'' is invariable. In the
autooscillatory modes, it bears the chaotic character.

The studies indicate that, in the interval $\mu _0 \in$ $\in (0.43,
0.438)$, the stable autoperiodic modes of the system are transformed
into chaotic ones.

\begin{figure*}%
\vskip1mm
\includegraphics[width=15.5cm]{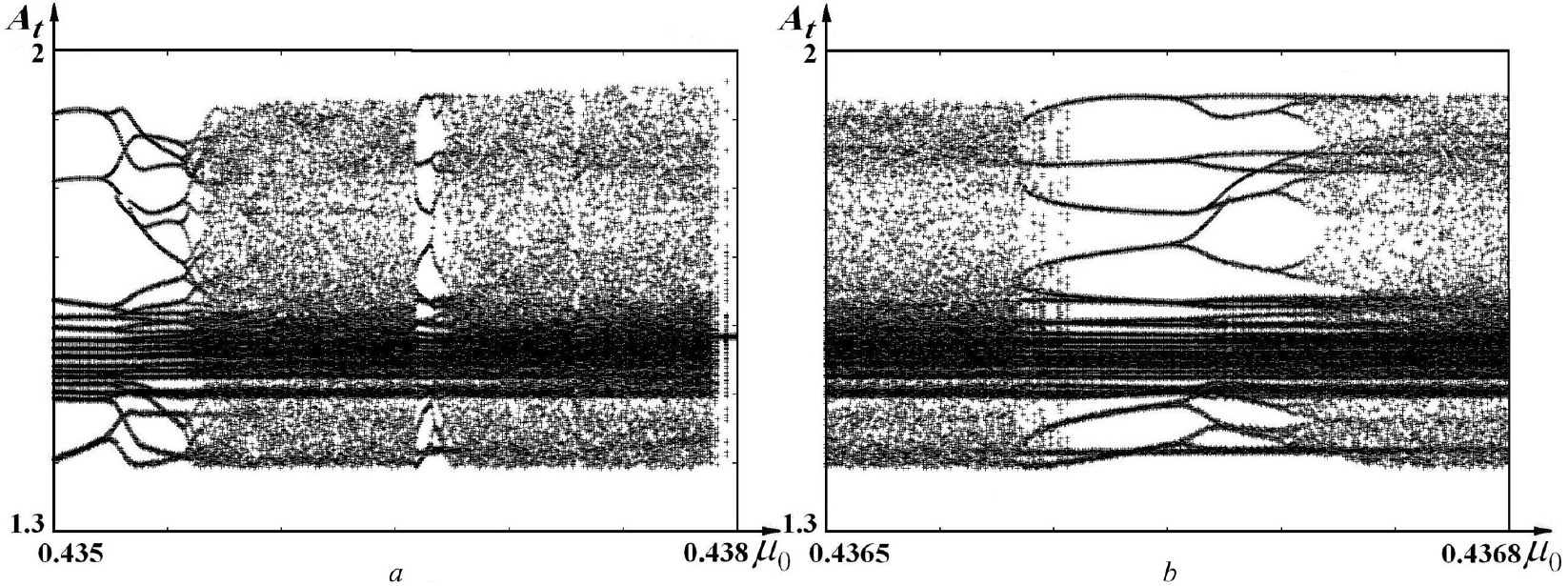}
\vskip-3mm\parbox{15.5cm}{\caption{Phase-parametric diagram of the
system for the variable $A_t (t)$: $\mu _0 \in (0.435, 0.438)$
(\textit{a}); $\mu _0 \in (0.4365, 0.4368)$~(\textit{b})\label{fig:3}}}
\end{figure*}
\begin{figure*}%
\vskip3mm
\includegraphics[width=15.5cm]{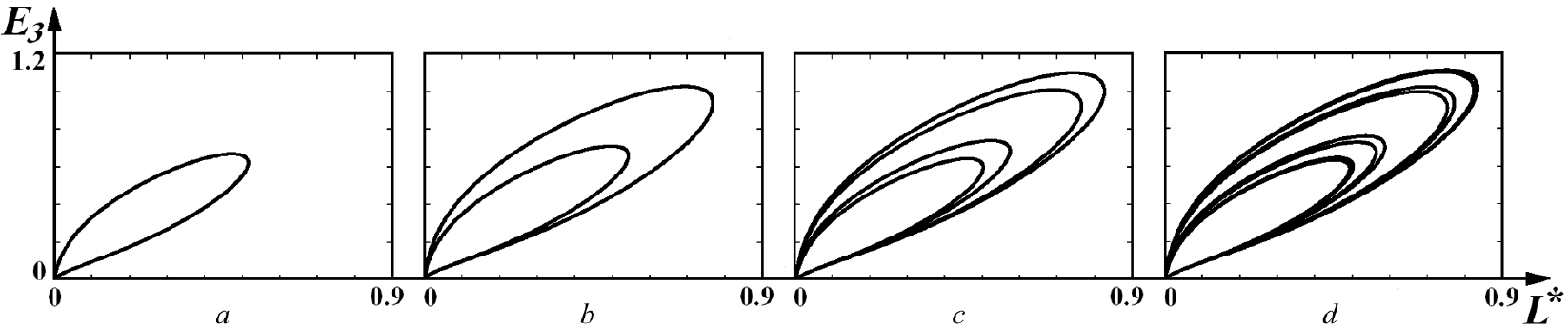}
\vskip-3mm\parbox{15.5cm}{\caption{Projections of the phase
portraits of regular attractors of the system in the plane $(L^\ast
, E_3 )$: $1\cdot 2^0$ for $\mu _0 =0.43$ (\textit{a}); $1\cdot 2^1$
for $\mu _0 =0.435$ (\textit{b}); $1\cdot 2^2$ for $\mu _0 =0.43555$
(\textit{c}); and $1\cdot 2^4$
for $\mu _0 =0.43563$ (\textit{d})\label{fig:4}}}
\end{figure*}

Let us study the dependence of the oscillatory dynamics of the
metabolic process of hemostasis on the dissipation of cholesterol
from a blood vessel determined by the parameter $\mu _0 $. In
Fig.~\ref{fig:3},~\textit{a},~\textit{b}, we show the
phase-parametric diagram of the system for the variable $A_t (t)$
under the change in $\mu _0 $ in the appropriate intervals. To
construct the phase-parametric diagrams, we used the method of
cutting. In the phase space of trajectories of the system, we place
the cutting plane for $E_2 =1.2$. As the trajectory crosses this
plane in some direction, we mark the value of chosen variable ($A_t
(t)$ in this case) on the phase-parametric diagram. Such choice is
explained by the symmetry of oscillations of enzyme $E_2 $ relative
to this point in the multiply calculated earlier modes. For every
value of $A_t (t)$, we mark the point of intersection of this plane
by the trajectory in one direction, when the trajectory approaches
an attractor. If a multiple periodic limiting cycle appears, the
points marked on the plane will coincide in a period. In the case of
the deterministic chaos, the points of intersection are located
chaotically.

The scenario of the transition from autoperiodic modes to chaotic
ones is presented in Fig.~\ref{fig:4}. The transition occurs due to
the doubling of a period. As $\mu _0 $ increases further, the
autoperiodic mode is destroyed after the 4-th bifurcation, and chaos
arises as a result of the intermittence.

\begin{figure*}%
\vskip1mm
\includegraphics[width=13cm]{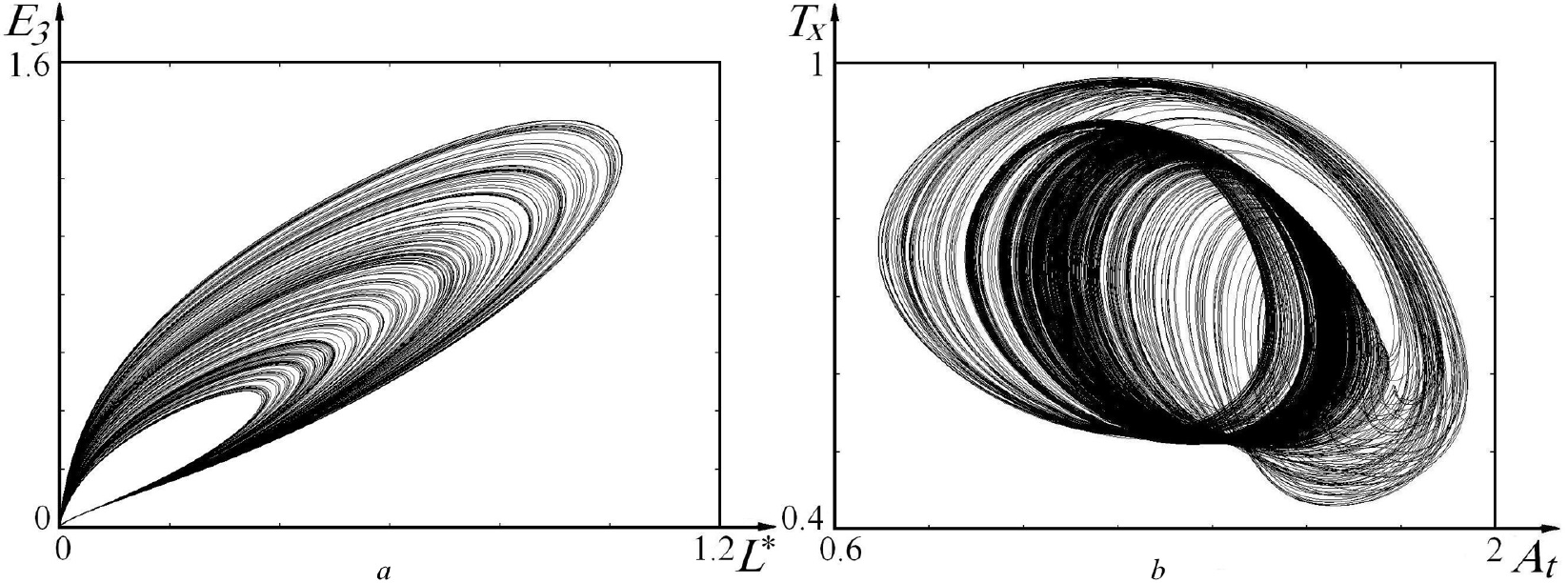}\\[2.5mm]
\includegraphics[width=13cm]{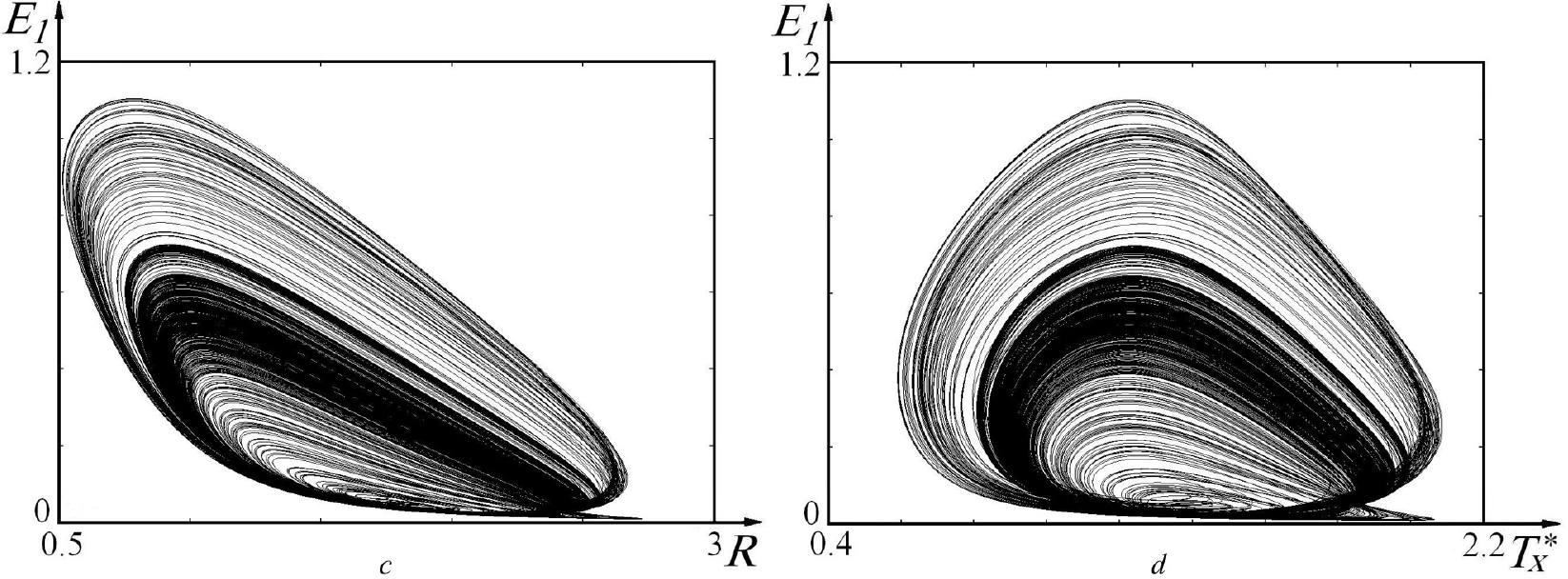}
\vskip-2mm\parbox{13.0cm}{\caption{Projections of phase portraits of
the strange attractor $2^x$ for $\mu _0 =0.437$: in the plane
$(L^\ast ,E_3 )$ (\textit{a}), in the plane $(A_t ,T_x )$
(\textit{b}), in the plane $(R,E_1 )$ (\textit{c}), and in the
plane $(T_x^\ast ,E_1 )$ (\textit{d})\label{fig:5}}}
\end{figure*}

It is seen from Fig.~\ref{fig:3},~\textit{b} that the periodicity
window arises between chaotic modes in the interval $\mu _0 \in
(0.43657, 0.43672)$. The scenario of the appearance of period
doubling bifurcations in it and the formation of chaotic modes is
analogous to that in Fig.~\ref{fig:4}. Some examples of the
projections of phase portraits of a chaotic attractor are presented
in Fig.~\ref{fig:5},~\textit{a, b, c, d}.
Figure~\ref{fig:5},~\textit{c,~d} indicates that the given strange
attractor is formed due to a funnel. In the funnel, the
trajectories, which approach one another in some directions and move
away in other directions, are mixed. Under a small fluctuation, the
periodic process becomes unstable, and the deterministic chaos
\mbox{arises.}\looseness=1

\begin{table*}[!]
\vskip7mm \noindent\caption{Lyapunov's indices, KS-entropy,
``predictability horizon'',\\ and Lyapunov's dimension of the
fractality of strange attractors calculated in different
modes}\vskip3mm\tabcolsep16.9pt

\noindent{\footnotesize\begin{tabular}{|l|c|c|c|c|c|c|c|}
 \hline \multicolumn{1}{|c}
{\rule{0pt}{5mm}$\mu _0 $} & \multicolumn{1}{|c}{Attractor}&
\multicolumn{1}{|c}{$\lambda _1 $}& \multicolumn{1}{|c}{$\lambda
_2$}& \multicolumn{1}{|c}{$\lambda _3$}& \multicolumn{1}{|c}{$h$}&
\multicolumn{1}{|c}{$t_{\min } $}&
\multicolumn{1}{|c|}{$D_{F_r}$}\\[2mm]%
\hline%
\rule{0pt}{5mm}0.43& $1\cdot 2^0$& .00003& --.00301& --.06362& --&
--&
-- \\
 0.43563& $1\cdot 2^4$& .00003& --.00282& --.00546& --& --&
-- \\
0.4365& $2^x$& .00170& ~\,.00014& --.00323& .00170& 588.24&
2.53 \\
0.4368& $2^x$& .00181& ~\,.00013& --00368& .00181& 552.25&
2.49 \\
0.437& $2^x$& .00192& ~\,.00020& --.00432& .00192& 520.83&
2.44 \\
0.4375& $2^x$& .00244& ~\,.00017& --.00310& .00244& 409.84&
2.79 \\
0.4378& $2^x$& .00263& ~\,.00020& --.00311& .00263& 380.23&
2.85 \\[2mm]%
\hline
\end{tabular}}\vspace*{2mm}
\end{table*}

While studying other modes of the phase-pa\-ra\-met\-ric disgrams in
Fig.~\ref{fig:3},~\textit{a,~b}, it is impossible in proper time to
determine the type of the attractor, to which a specific mode can be
referred. It can be a multiple stable or quasistable autoperiodic
cycle, as well as a strange attractor.

For the unambiguous identification of the type of obtained
attractors and for the determination of their stability for the
chosen parameters, we calculated the full spectra of Lyapunov's
indices $\lambda _1 ,\lambda _2 ,...,\lambda _{12} $ and their sum
$\Lambda =\sum _{j=1}^{12} \lambda _j $ by Benettin's algorithm with
orthogonalization of the perturbation vectors within the
Gram--Schmidt method \cite{18}.

As an example for comparison, we present some results of
calculations of the full spectra of Lyapunov's indices in Table~1.
To decrease table's size, we considered only the first three indices
$\lambda _1 -\lambda _3 $. The values of $\lambda _4 -\lambda _{12}
$ and $\Lambda $ are omitted, because they are not significant for
the presentation of our results. The obtained numbers are rounded to
the fifth decimal point. Based on the data in Table~1, we have
calculated some other indices for strange attractors.

With the use of Pesin's theorem \cite{45}, we calculated the
KS-entropy (Kolmogorov--Sinai entropy) and Lyapunov's indicator
``predictability horizon'' \cite{46}. Lyapunov's dimension of the
fractality of strange attractors was found by the Kaplan--Yorke
formula \cite{47,48}:

By the calculated indicators, we may judge about the difference of
geometric structures of the given strange attractors.

Having calculated successively various strange attractors, we can
indicate some regularity in the hierarchy of their chaotic behavior.
According to changes in the indices, the geometric shape of
attractors of the system varies.

The autooscillations in the metabolic process of hemostasis of a
blood vessel arise due to the interaction of two systems of blood
(thrombosis and antithrombosis), which is regulated by the
concentration of cyclic adenosine monophosphate. The presence of
``bad cholesterol'' in blood causes the desynchronization of these
systems. In this case, the chaotic modes appear in the metabolism of
hemostasis. LDLs affect the binding of thrombocytes and are
deposited on the walls of vessels. This induces the autocatalysis of
cholesterol in blood.

Thus, as the amount of cholesterol in blood chan\-ges, the
hemostasis of blood vessels adapts to this change, by conserving its
functionality in this case.

\section{Conclusions}

Within a mathematical model, we have studied the influence of ``bad
cholesterol'' on the metabolic process of hemostasis of blood
vessels. The system with metabolic process is considered as a
dissipative system, whose input substances are arachidonic acid and
low-density lipoproteins (LDLs). The interaction between the systems
of thrombosis and antithrombosis leads to the appearance of
stationary or autoperiodic modes. The presence of ``bad
cholesterol'' breaks the balance of this process. It is the
autocatalyst of the level of LDLs in blood. The interval of the
intensity of dissipation of cholesterol from blood, in which the
chaotic autooscillations in the metabolism of hemostasis can arise,
is determined. The phase-parametric diagrams of autooscillatory
modes, which depend on the dissipation of cholesterol from blood,
are constructed. We have found the scenario of period doubling
bifurcations existing until the aperiodic modes of strange
attractors arise as a result of the intermittence. We have
calculated the strange attractors, which appear due to the formation
of the mixing funnel, and the full spectra of Lyapunov's indices for
various modes. For the strange attractors, the KS-entropies,
``predictability horizons'', and Lyapunov's dimensions of the
fractality of attractors are determined. It is shown that the reason
for changes in the metabolic process of hemostasis of blood vessels
can be an insufficient intensity of dissipation of cholesterol from
blood. The obtained results allow one to study the influence of
low-density lipoproteins on the self-organization of the metabolic
process of hemostasis of blood vessels and the development of
stenosis. The reasons for changes of a physiological state of a
blood vessel leading to thrombophilia are found. The influence of
statins and other physiologically active substances on a decrease in
cholesterol in blood vessels is estimated.

\vskip3mm \textit{The work is supported by the project
No.\,0113U001093 of the National Academy of Sciences of Ukraine.}

\rezume{%
В.Й.\,Грицай}{САМООРГАНІЗАЦІЯ\\ ТА ХАОС
 В МЕТАБОЛІЗМІ ГЕМОСТАЗУ\\ У КРОВОНОСНІЙ СУДИНІ} {В даній роботі побудована математична модель метаболічного
процесу становлення гемостазу в кровоносній судині. На відміну від
раніш побудованої поліферментної простациклін-тромбоксанової системи
крові, в цій моделі вперше враховано вплив на гемостаз рівня
``поганого холестерина''~-- ліпопротеїдів низької щільності (ЛПНЩ).
Знайдено умови, при яких виникає самоорганізація системи,
утворюються режими автоколивань та хаосу в метаболічному процесі, що
впливає на становлення гемостазу та виникнення тромбофілії. За
допомогою фазопараметричної діаграми досліджено сценарій їх
виникнення. Знайдено біфуркації подвоєння періоду та переходу до
хаотичних коливань внаслідок перемєжаємості. Отримані дивні
аттрактори утворюються внаслідок воронки перемішування. Розраховані
їх повні спектри показників Ляпунова, КС-ентропії, ``горизонти
передбачуваності'' і ляпуновські розмірності дивних аттракторів.
Зроблено висновки про причи зміни циклічності в даному метаболічному
процесі, його стійкості і фізіологічному прояві в кровоносній
системі. Оцінено роль фізіологічно-активних речовин для зменшення
рівня холестерина в судинах.}

\vspace*{-3mm}
\rezume{%
В.И.\,Грицай}{ САМООРГАНІЗАЦИЯ\\ И ХАОС
 В МЕТАБОЛИЗМЕ ГЕМОСТАЗА\\ В КРОВЕНОСНОМ СОСУДЕ} {В данной работе построена математическая модель метаболического
процесса становления гемостаза в кровеносном сосуде. В отличие от
ранее построеной полиферментной простациклин-тромбоксановой системы
крови, в данной модели впервые учтено влияние на гемостаз уровня
``плохого холестерина''~-- липопротеидов низькой плотности (ЛПНП).
Найдены условия, при которых возникает самоорганизация системы,
образуются режимы автоколебаний и хаоса в метаболическом процессе,
что влияет на становление гемостаза и возникновение тромбофилии. При
помощи фазопараметрической диаграммы исследовано сценарій их
возникновения. Найдены бифуркации удвоения периода и перехода к
хаотическим колебаниям вследствие перемежаємості. Полученные
странные аттракторы возникают вследствие воронки перемешивания.
Рассчитаны их полные спектры показателей Ляпунова, КС-энтропии,
``горизонты предсказуемости'' и ляпуновские размерности странных
аттракторов. Сделаны выводы о причинах изменения цикличности в
данном метаболическом процессе, его устойчивости и физиологическом
проявлении в кровеносной системе. Оценена роль
физиологически-активных веществ для уменьшения уровня холестерина в
сосудах.}

\end{document}